\documentclass[a4paper,12pt]{article}
\usepackage{amsmath,amsfonts,amssymb}
\usepackage{graphicx}

\begin{document}

\begin{center}
{\LARGE\bf Linear radial Regge trajectories for mesons with any
quark flavor\footnote{A talk presented at QUARKS-2016.}}
\end{center}
\bigskip
\begin{center}
{\large S. S. Afonin and I. V. Pusenkov}
\end{center}

\begin{center}
{\small Saint Petersburg State University, 7/9 Universitetskaya
nab., St.Petersburg, 199034, Russia}
\end{center}
\bigskip

\begin{abstract}
In the Regge phenomenology, the radial spectrum of light mesons is
given by a linear relation $M_n^2=a(n+b)$, where $a$ is a
universal slope, the dimensionless intercept $b$ depends on
quantum numbers, and $n$ enumerates the excited states in radial
recurrences. The usual extensions of this relation to heavy
quarkonia in the framework of hadron string models typically lead
to strong nonlinearities which seem to be at variance with the
available experimental data. Introducing a radially static string
picture of mesons, we put forward a linear generalization
$(M_n-m_1-m_2)^2=a(n+b)$, where $m_{1,2}$ are quark masses. The
vector channel contains enough experimental states to check this
new relation and a good agreement is observed. It is shown that
this generalization leads to a simple estimate of current quark
masses from the radial spectra.
\end{abstract}

\bigskip

The Regge phenomenology grew out of the dual
amplitudes~\cite{collins} and gave rise to various hadron string
models. This approach continues to play an important role in the
study of hadron spectroscopy. The dual amplitudes and some related
string approaches predict the following behavior of meson masses,
\begin{equation}
\label{1}
M_n^2=a(J+n+c), \qquad J,n=0,1,2,\dots,
\end{equation}
where $J$ is the spin, $n$ enumerates the radially excited states,
$a$ represents a universal slope and $c$ is a constant. The
relation~\eqref{1} reproduces the classical Regge behavior
$M^2\sim J$ observed in the light baryons and mesons. Also this
relation predicts the equidistant daughter Regge trajectories. The
number $n$ enumerates the states on these "radial" Regge
trajectories. The available experimental data on light non-strange
mesons seem to confirm the linear Regge
behavior~\eqref{1}~\cite{ani,klempt,cl1,cl2}.

Various attempts to generalise the relation~\eqref{1} to the
sector of heavy quarks resulted in the emergence of strong
non-linearities with respect to $J$ and $n$. Let us look, however,
at the experimental data. A relatively rich set of data on radial
recurrences in this sector exists only for the unflavored vector
heavy quarkonia. Using the relevant data from the Particle
Data~\cite{pdg} which are displayed in Table~1 (see Ref.~\cite{AP}
for detailed discussions on the data choice) we plot the masses
squared of known $\omega$, $\varphi$, $\psi$, and
$\Upsilon$-mesons as a function of consecutive number $n$ in
Figs.~1 -- 4.
\begin{table}
\caption{\small The masses of known $\omega$, $\phi$, $\psi$ and
$\Upsilon$ mesons (in MeV) which are used in our
analysis. The experimental error is not displayed if it is less
than 1~MeV.}
\begin{center}
{
\begin{tabular}{|c|ccccc|}
\hline
$n$ & $0$ & $1$ & $2$ & $3$ & $4$\\
\hline
$M_\omega$ & $783$ & $1425 \pm 25$ & $1670 \pm 30$ & $1960 \pm 25$ & $2205 \pm 30$ \\
$M_\phi$ & $1020$ & $1680 \pm 20$ & --- & $2175 \pm 15$ & --- \\
$M_\psi$ & $3097$ & $3686$ & $4039 \pm 1$ & $4421 \pm 4$ & --- \\
$M_\Upsilon$ & $9460$ & $10023$ & $10355$ & $10579 \pm 1$ & $10865 \pm 8$ \\
\hline
\end{tabular}}
\end{center}
\end{table}
\begin{figure}[htb]
\center{\includegraphics[width=0.5\linewidth]{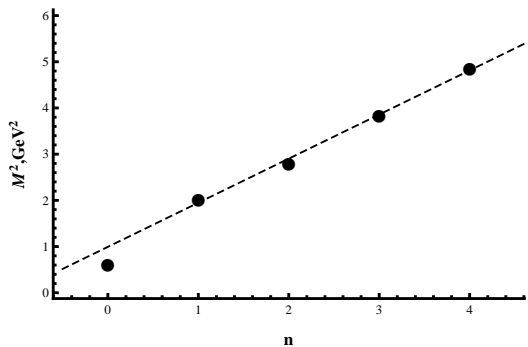}}
\vspace{-0.3cm}
\caption{The spectrum of $\omega$-mesons.
    The experimental points (for this and subsequent figures) are taken from Table 1.}
\end{figure}
\begin{figure}[htb]
\center{\includegraphics[width=0.5\linewidth]{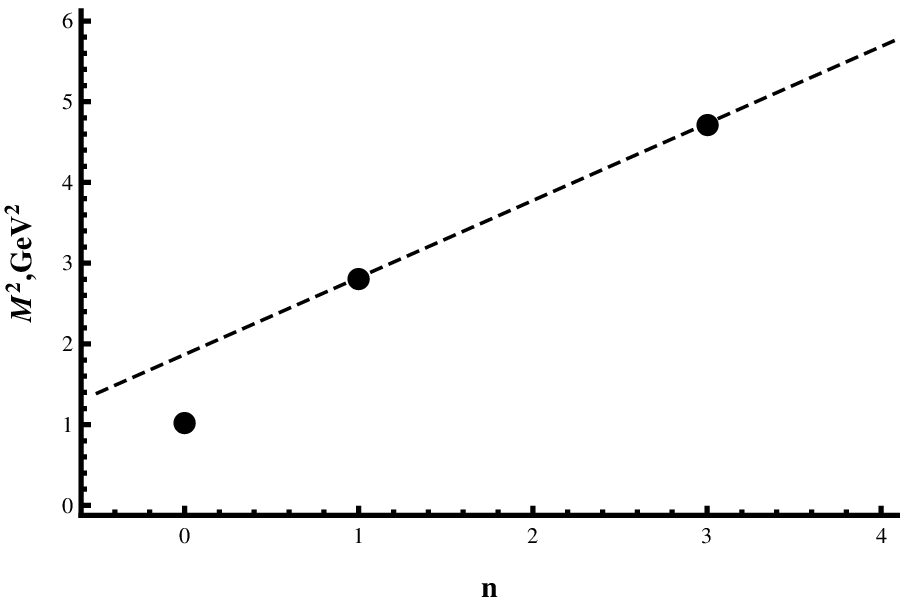}}
\vspace{-0.3cm}
\caption{The spectrum of $\phi$-mesons.}
\end{figure}
\begin{figure}[htb]
\center{\includegraphics[width=0.5\linewidth]{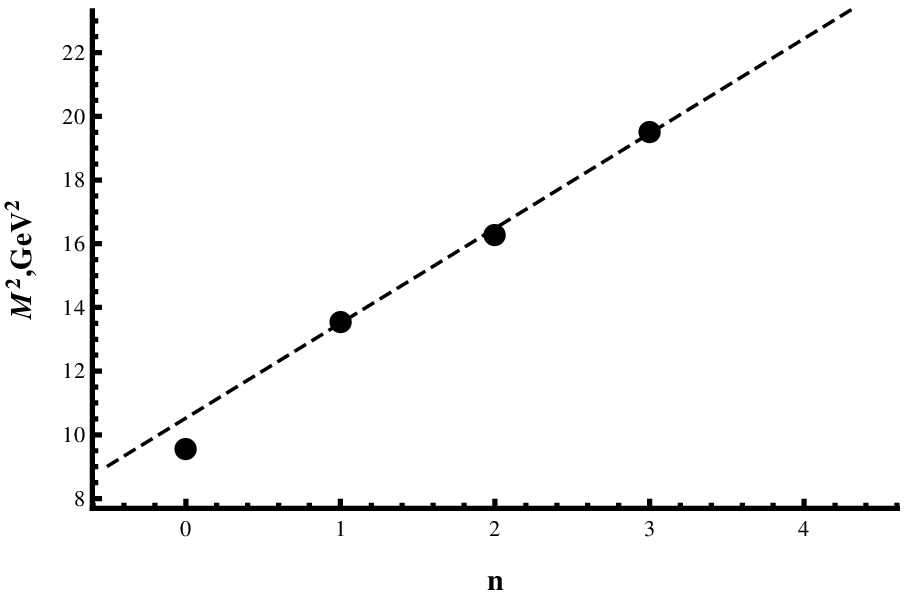}}
\vspace{-0.3cm}
\caption{The spectrum of $\psi$-mesons.}
\end{figure}
\begin{figure}[htb]
\center{\includegraphics[width=0.5\linewidth]{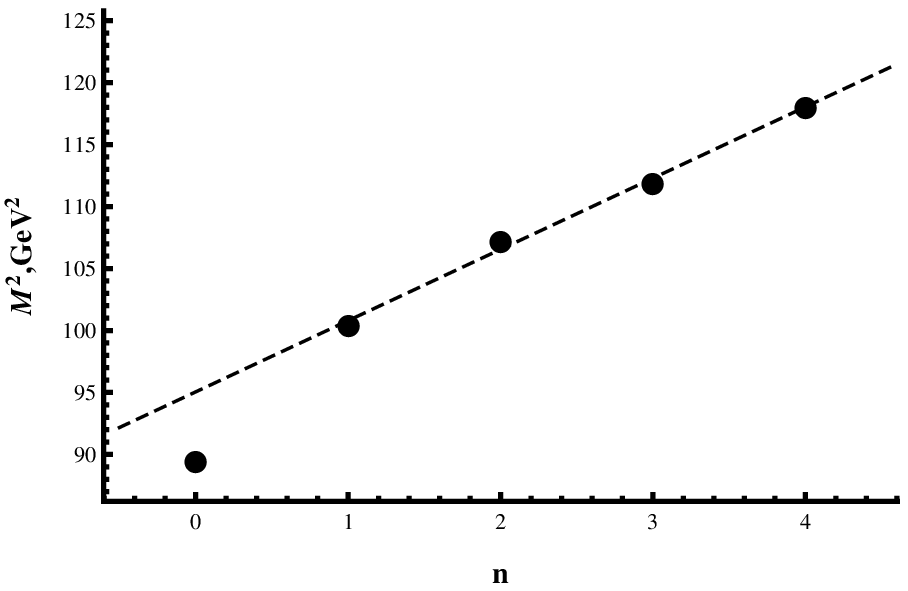}}
\vspace{-0.3cm}
\caption{The spectrum of $\Upsilon$-mesons.}
\end{figure}
Aside from the ground state, the masses approximately lie on the
linear radial trajectories
\begin{equation}
\label{2}
M_n^2=a(n+b),
\end{equation}
where we re-denoted the vector intercept $b=1+c$. The slope and
intercept in~\eqref{2} depend strongly on the quark flavor, see
Table~2.
\begin{table}
\caption{\small The radial Regge trajectories~\eqref{2} (in
GeV$^2$) for the data from Table 1 (see text).}
\begin{center}
{
\begin{tabular}{|c|cc|}
\hline
$M_n^2$ & Fit (a) & Fit (b)\\
\hline
$M_\omega^2$ & $1.03 (n+0.74)$ & $0.95 (n+1.04)$\\
$M_\phi^2$ & $1.19 (n+1.07)$ & $0.95 (n+1.96)$\\
$M_\psi^2$ & $3.26 (n+3.03)$ & $2.98 (n+3.53)$\\
$M_\Upsilon^2$ & $6.86 (n+11.37)$ & $5.75 (n+16.54)$\\
\hline
\end{tabular}}
\end{center}
\end{table}

The universal linearity seen in Figs.~1 -- 4 suggests that some
universal gluodynamics lies behind the observed behavior. We
propose a simple string scheme explaining this universality. The
central idea is that at the conditions when a quark-antiquark pair
form a resonance in the QCD vacuum, the pair can be viewed as a
radially static system. The binding is provided by the exchange of
some massless particle (the pion or gluon in concrete realizations
of the scheme). And one must quantize the motion of this particle
(not the radial motion of quarks as in the standard hadron string
approaches!). The total energy (mass) of the system is
\begin{equation}
\label{3}
M=m_1+m_2+p+\sigma r.
\end{equation}
Here $m_1$ and $m_2$ are the masses of quark and antiquark
separated by the distance $r$, $p$ is the momentum of exchanged
particle, and $\sigma$ represents the standard string tension.
Let us apply the semiclassical quantization to the momentum $p$
\begin{equation}
\label{4}
\int_0^l p\,dr=\pi(n+b),\qquad n=0,1,2,\dots.
\end{equation}
Here $l$ is the maximal quark separation and the constant $b$
depends on the boundary conditions. Substituting $p$
from~\eqref{3} to~\eqref{4} and making use of the definition
$\sigma=\frac{M}{l}$ we obtain the linear radial trajectory
\begin{equation}
\label{5}
(M_n-m_1-m_2)^2=2\pi\sigma(n+b).
\end{equation}

In our unflavored case $m_1=m_2\equiv m$ and the relation~\eqref{5}
can be simplified to
\begin{equation}
\label{6}
(M_n-2m)^2=a(n+b),
\end{equation}
which is our generalization of the linear spectrum~\eqref{2} to
non-zero quark masses. In this relation, the universal
gluodynamics (the slope $a$) and dependence on quantum numbers
(the dimensionless intercept $b$) are clearly separated from the
contribution of quark masses. For this reason, the parameters $a$
and $b$ in the relation~\eqref{6} should be flavor-independent.

Let us test our expectations. We will consider two cases --- with
the light quark mass set to zero (Fit~I) and with all quark masses
unfixed (Fit~II). The results of interpolation of the data in
Table~1 by the ansatz~\eqref{6} are given in Table~3. The ensuing
two variants for the spectrum are depicted in Fig.~5 and Fig.~6.
Taking into account a simplicity of the relation~\eqref{6}, the
agreement is remarkable.
\begin{table}
\caption{\small The quark masses (in GeV), the slope $a$ (in
GeV$^2$) and the dimensionless intercept parameter $b$ in the
relation~\eqref{6}.}
\begin{center}
{\footnotesize
\begin{tabular}{|c|cc|}
\hline
 & Fit I & Fit II \\
\hline
$m_{u,d}$ & 0 & 0.36\\
$m_s$ & 0.13 & 0.49 \\
$m_c$ & 1.17 & 1.55\\
$m_b$ & 4.33 & 4.69\\
\hline
$a$ & 1.10 & 0.49\\
\hline
$b$ & 0.57 & 0.00\\
\hline
\end{tabular}}
\end{center}
\end{table}
\begin{figure}[htb]
\center{\includegraphics[width=0.7\linewidth]{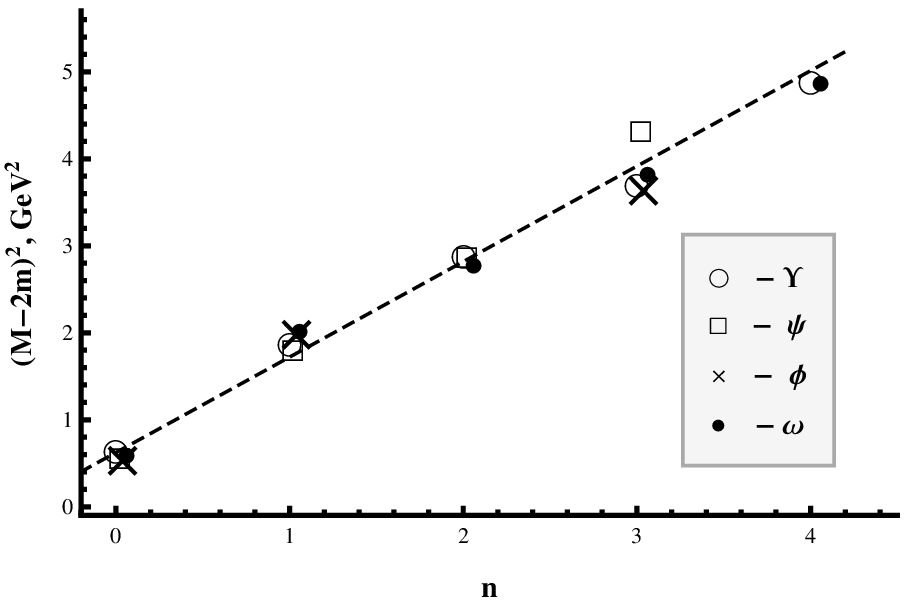}}
\vspace{-0.3cm}
\caption{The spectrum~\eqref{6} for $m_{u,d}$ fixed (Fit~I).}
\end{figure}
\begin{figure}[htb]
\center{\includegraphics[width=0.7\linewidth]{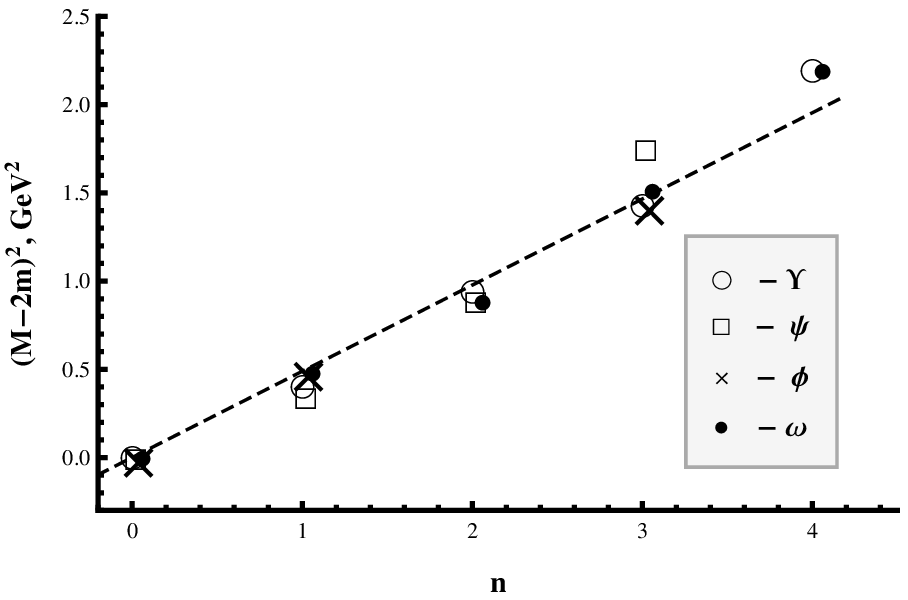}}
\vspace{-0.3cm}
\caption{The spectrum~\eqref{6} for $m_{u,d}$ unfixed (Fit~II).}
\end{figure}

The results in Table~3 demonstrate that the radial meson
trajectories are able to "measure" the current quark masses with
surprisingly good accuracy. If we set $m_{u,d}=0$, the current
masses of other quarks turn out to be very close to their
phenomenological values~\cite{pdg}. If we keep all quark masses
unfixed, they acquire an additional contribution about $360$ MeV.
This contribution may be interpreted as an averaged value of
constituent quark mass emerging due to the chiral symmetry
breaking in QCD. A more detailed analysis (see Ref.~\cite{AP})
shows that the Fit~I works better and the interpretation with the
current quark masses looks preferable. Many related discussions
and further fits can be found in Ref.~\cite{AP}. An independent
fit of the relation~\eqref{6} was also performed by the authors of
Ref.~\cite{univer_arriola}.

In summary, we proposed a new generalization of linear radial
Regge trajectories to the case of massive quarks. Within this
generalization, the form of contribution to the meson masses due
to confinement is universal for any quarkonia. Our generalization
is well consistent with the experimental data on the unflavored
vector mesons. Although our considerations were simple and did not
take into account various effects which may cause some mass
shifts, the quality of final fits is comparable with typical
results of semirelativistic potential models, hadron strings and
other technically nontrivial approaches.

A natural question emerging after our analysis is whether the
relation~\eqref{6} can be extended to other types of mesons?
Unfortunately, the available experimental data are too scarce for
making any definite conclusion. A possible extensions of~\eqref{6}
is
\begin{equation}
\label{10}
(M_n-m_1-m_2)^2=a(n+x+b),
\end{equation}
where we may have $x=\beta L$ or $x=\beta J$ (here $L$ and $J$
mean the orbital momentum of valent quarks and total spin). The
constant $\beta$ should be fixed from the phenomenology. An
intriguing possibility $\beta=1$ would lead to a large degeneracy
observed in the light non-strange mesons~\cite{klempt,cl1,cl2}.
The fact that the slope $a=2\pi\sigma$ in~\eqref{5} coincides with
the slope of rotating open string may give a theoretical
explanation for this degeneracy.

\underline{\it Acknowledgments.} The work was supported by the
Saint Petersburg State University research grant 11.38.189.2014
and by the RFBR grant 16-02-00348-a.


\begin{thebibliography}{99}

\bibitem{collins} P.D.B. Collins, Phys. Rept. C {\bf 1}, 103 (1971).
\bibitem{ani} A.~V.~Anisovich, V.~V.~Anisovich and A.~V.~Sarantsev,
Phys. Rev. D~{\bf 62}, 051502(R) (2000); D.~V.~Bugg, Phys. Rept.
{\bf 397}, 257 (2004).
\bibitem{klempt} E. Klempt and A. Zaitsev, Phys. Rep. {\bf 454}, 1 (2007).
\bibitem{cl1} S.~S.~Afonin, Phys. Lett. B {\bf 639}, 258 (2006);
Eur. Phys. J. A {\bf 29}, 327 (2006).
\bibitem{cl2} S.~S.~Afonin, Phys. Rev. C {\bf 76}, 015202 (2007);
Mod. Phys. Lett. A {\bf 22}, 1359 (2007); M. Shifman and A.
Vainshtein, Phys. Rev. D {\bf 77}, 034002 (2008).
\bibitem{pdg} K. A. Olive {\it et al.} (Particle Data Group), Chin. Phys. C
{\bf 38}, 090001 (2014).
\bibitem{AP} S.~S.~Afonin and I.~V.~Pusenkov,
Phys. Rev. D {\bf 90}, 094020 (2014); Mod. Phys. Lett. A {\bf 29},
1450193 (2014).
\bibitem{univer_arriola} P.~Masjuan, E.~R.~Arriola and
W.~Broniowski, EPJ Web Conf.\ {\bf 73}, 04021 (2014).
\end{thebibliography}
\end{document}